\documentclass[sigconf]{acmart}

\usepackage{booktabs} 
\usepackage{wrapfig}
\usepackage{graphicx}
\usepackage[ruled]{algorithm2e} 
\usepackage{array}
\usepackage{multirow}

\SetAlFnt{\small}
\SetAlCapFnt{\small}
\SetAlCapNameFnt{\small}
\SetAlCapHSkip{0pt}

\copyrightyear{2024}
\acmYear{2024}
\setcopyright{acmlicensed}\acmConference[SIGGRAPH Conference Papers '24]{Special Interest Group on Computer Graphics and Interactive Techniques Conference Conference Papers '24}{July 27-August 1, 2024}{Denver, CO, USA}
\acmBooktitle{Special Interest Group on Computer Graphics and Interactive Techniques Conference Conference Papers '24 (SIGGRAPH Conference Papers '24), July 27-August 1, 2024, Denver, CO, USA}
\acmDOI{10.1145/3641519.3657401}
\acmISBN{979-8-4007-0525-0/24/07}

\usepackage{xspace}
\makeatletter
\DeclareRobustCommand\onedot{\futurelet\@let@token\@onedot}
\def\@onedot{\ifx\@let@token.\else.\null\fi\xspace}

\makeatother

\newcommand{\bH}{\mathbf{H}}

\newcommand{\bJ}{\mathbf{J}}

\newcommand{\bM}{\mathbf{M}}

\newcommand{\bS}{\mathbf{S}}
\newcommand{\bT}{\mathbf{T}}

\newcommand{\bX}{\mathbf{X}}

\newcommand{\bc}{\mathbf{c}}

\newcommand{\be}{\mathbf{e}}
\newcommand{\bff}{\mathbf{f}}

\newcommand{\bu}{\mathbf{u}}

\newcommand{\bx}{\mathbf{x}}

\newcommand{\tran}{\mathsf{T}}

\newcommand{\restcurvature}{\bar{\boldsymbol{\kappa}}}
\newcommand{\curvature}{\boldsymbol{\kappa}}

\usepackage{amsmath}

\DeclareMathOperator*{\argmin}{arg\,min}

\begin{document}
\title{Modal Folding: Discovering Smooth Folding Patterns for Sheet Materials using Strain-Space Modes}


\author{Pengbin Tang}
\affiliation{
  \institution{ETH Z{\"u}rich}
  \country{Switzerland}
}
\affiliation{
  \institution{Universit\'e de Montr\'eal}
  \country{Canada}
}
\email{petang@ethz.ch}

\author{Ronan Hinchet}
\affiliation{
  \institution{ETH Z{\"u}rich}
  \country{Switzerland}
}
\email{ronan.hinchet@srl.ethz.ch}

\author{Roi Poranne}
\affiliation{
  \institution{ETH Z{\"u}rich}
  \country{Switzerland}
}
\affiliation{
  \institution{University of Haifa}
  \country{Israel}
}
\email{roi.poranne@inf.ethz.ch}

\author{Bernhard Thomaszewski}
\affiliation{
  \institution{ETH Z{\"u}rich}
  \country{Switzerland}
}
\email{bthomasz@inf.ethz.ch}

\author{Stelian Coros}
\affiliation{
  \institution{ETH Z{\"u}rich}
  \country{Switzerland}
}
\email{stelian@inf.ethz.ch}

\begin{abstract}

Folding can transform mundane objects such as napkins into stunning works of art.
However, finding new folding transformations for sheet materials is a challenging problem that requires expertise and real-world experimentation.
In this paper, we present Modal Folding---an automated approach for discovering energetically optimal folding transformations, i.e., large deformations that require little mechanical work.
For small deformations, minimizing internal energy for fixed displacement magnitudes leads to the well-known elastic eigenmodes. While linear modes provide promising directions for bending, they cannot capture the rotational motion required for folding. To overcome this limitation, we introduce strain-space modes---nonlinear analogues of elastic eigenmodes that operate on per-element curvatures instead of vertices.
Using strain-space modes to determine target curvatures for bending elements, we can generate complex nonlinear folding motions by simply minimizing the sheet's internal energy.
Our modal folding approach offers a systematic and automated way to create complex designs.
We demonstrate the effectiveness of our method with simulation results for a range of shapes and materials, and validate our designs with physical prototypes.
\end{abstract}
%



\maketitle
\begin{figure}[h]
    \centering
    \includegraphics[width=\linewidth]{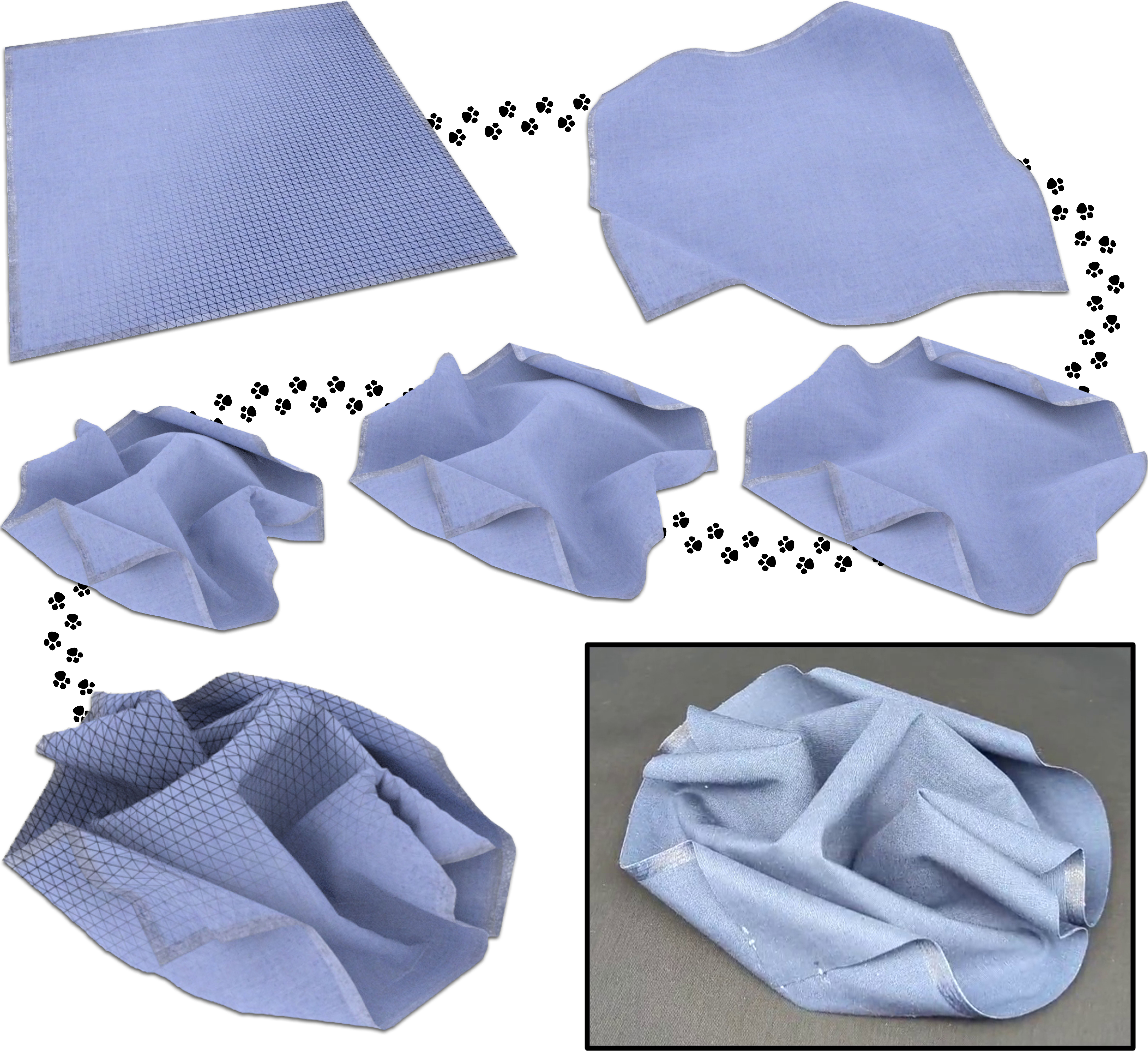}
    \caption{Our method automatically generates a diverse set of folding patterns using strain-space modes. Here we show a particular folding transformation for a square sheet of fabric and its physical prototype (\textit{bottom right}).}
    \label{fig:teaser}
\end{figure}

\begin{figure}[t]
    \centering
    \includegraphics[width=0.9\linewidth]{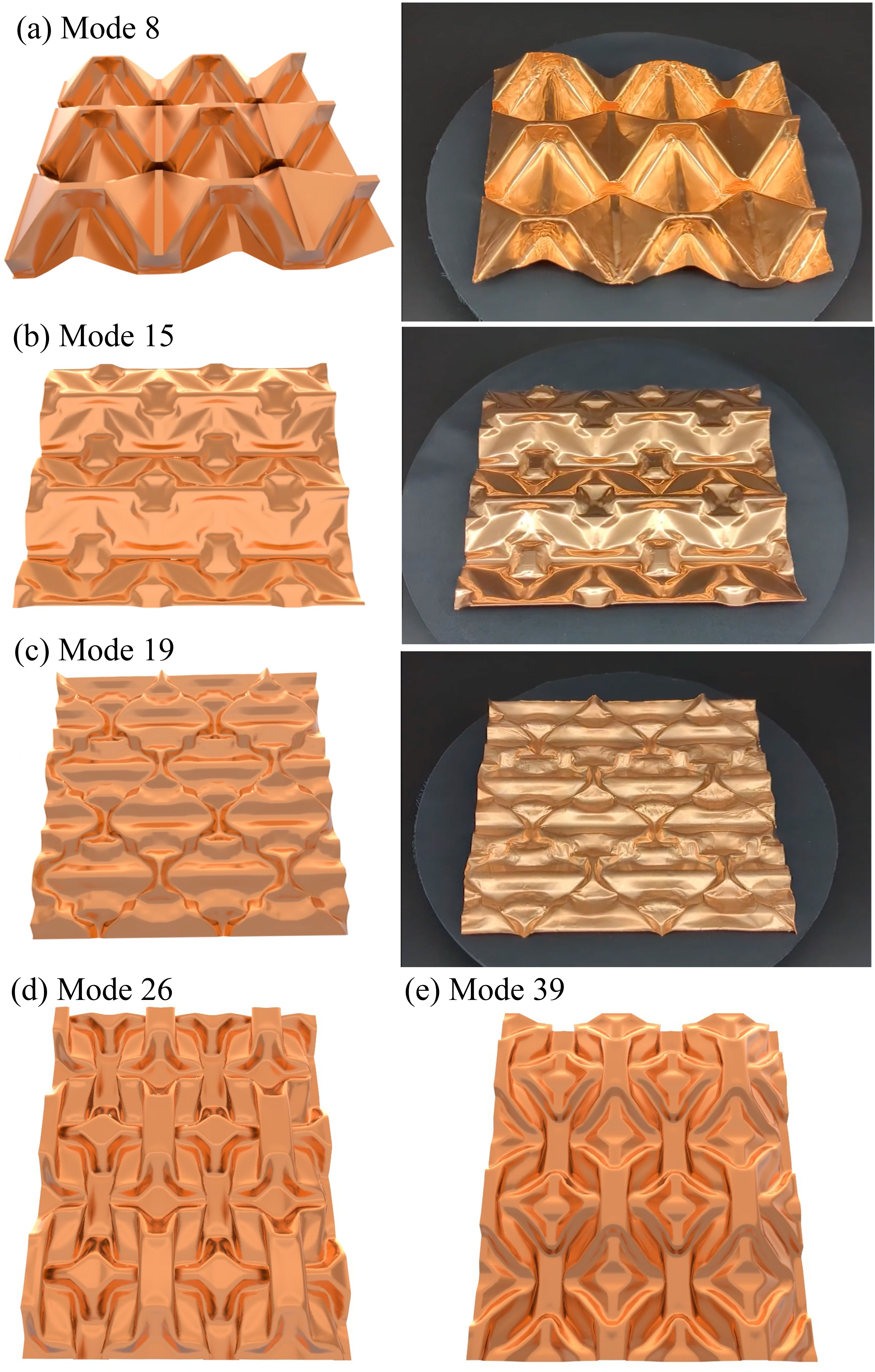}
    \caption{Periodic folding patterns obtained using Modal Folding (a-e) with reflection boundary conditions and their corresponding physical prototypes (\textit{right}) manufactured using copper sheets and 3D-printed press dies.}
    \label{fig:reflection}
\end{figure}

\section{Introduction}

In the hands of a skilled artist, a plain sheet of fabric can fold into a work of perplexing beauty. Finding such folding transformations, however, is an art in itself that has been curiously elusive to science. 
Nevertheless, systematic approaches for generating folding patterns would have wide implications beyond art, including deployable shading systems, architectural facade panels, and textile metamaterials.

In this work, we present Modal Folding---a novel approach for automatically discovering folding patterns of sheet materials.
Intuitively, folding transformations correspond to large deformations that require little mechanical work.
Informally, we can ask \textit{given a prescribed deformation magnitude, what is the configuration that minimizes the internal energy}?
Formulating this question as a constrained optimization problem reveals direct ties with the well-known elastic eigenmodes. Indeed, for linear materials and small deformations, elastic eigenmodes correspond to locally optimal solutions.
However, while linear modes point toward displacements that prioritize bending over stretching, they are inherently unable to capture the rotational motion required for folding. 

To address this limitation, we introduce \emph{strain-space modes} as a new nonlinear analogue of elastic eigenmodes.
Instead of following linear modes in position space, we consider their effect on the sheet's \textit{extrinsic deformation}. In the discrete setting, this gives rise to a set of per-element bending strains.
Collectively, we interpret these per-element strains as directions in a nonlinear strain space. Any strain-space configuration along such a direction can be mapped back to position space through closest-point projection, using the sheet's elastic energy as the metric.
Our key observation is that this approach---linear modes in strain space combined with energy-minimizing reconstruction--- gives rise to complex nonlinear folding transformations in position-space.
We leverage this insight to explore folding patterns by tracking strain-space modes in a physics-based and material-aware way.
Modal Folding offers a systematic and automated way to discover complex designs.
We demonstrate folding transformations for various input shapes, analyze the impact of material properties, and explore different types of periodic tilings. To illustrate the feasibility of our simulation results, we build physical prototypes using fabric, paper, and copper sheets.

\section{Related Work}

\paragraph{Folding Thin Sheets}

The problem of folding thin sheet materials has been studied intensively.
A central challenge lies in capturing sharp creases that can emerge during folding \cite{Schreck2016Nonsmooth}, which may require adapting mesh resolution and edge orientations \cite{Narain2012Adaptive}. 

Origami, the ancient art of paper folding, continues to inspire applications in  engineering \cite{meloni2021engineering,Callens2018}.
While traditional Origami designs are made from straight creases, curved folding opens up an even larger and fascinating design space \cite{Kilian2008Curved, Kilian2017String,Rabinovich2019Modeling,Jiang2019curved,Mundilova2019Mathematical,Tahouni2020SelfShaping}. 
Despite the prevalence of folding with sharp creases, smooth folding techniques also enjoy a multitude of applications, including robotic manipulation of clothes \cite{Li15Folding,DeGusseme22Effective}. Zhu et al. \shortcite{Zhu2013Soft} proposed a computational approach to soft folding that draws on principles from Origami. Folding can also be used to modulate the mechanical properties of fabrics. In this context, Ren et al.~\shortcite{Ren24Digital} introduced a parametric design tool for creating so-called smocking patterns.
Our approach likewise falls into the category of soft folding. However, instead of relying on heuristics or user-provided guidance, our method automatically discovers smooth folding transformation from a single input shape.

\paragraph{Eigenmodes in Animation and Design}
Eigenmodes have been used extensively in animation, especially in the context of subspace simulation. While linear modes are ideal for vibration analysis with small amplitude oscillations, they are notorious for generating artifacts for larger deformations. Numerous approaches have been proposed to improve upon linear modes, including modal derivatives \cite{Barbic2005Realtime}, higher-order descent directions \cite{Hildebrandt2011Interactive}, and correction of linearization artifacts \cite{Barbic2012Interactive,Choi2005Modal,Pan2015Subspace,Huang2011Interactive}. Besides subspace simulation, linear modes have also been used to generate fracture animations \cite{sellan2023breaking}.

Linear eigenmodes have also been widely used to design physical artifacts that involve small-amplitude vibrations, including metallophones \cite{umetani_nobuyuki_2010_1177917, Bharaj2015Computational}, wind instruments \cite{Umetani2016Printone}, and acoustic filters \cite{Li2016Acoustic}. Zhou et al. \shortcite{zhou2013Worst} and Liu et al.\shortcite{liu2022worst} leveraged linear eigenmodes to analyze and improve the stability of 3D printed designs. Similar in spirit, Zehnder et al. \shortcite{Zehnder2016Designing} employed sparse linear eigenanalysis to detect unwanted structural flexibility in curve networks.

While the aforementioned methods focus on small displacements, Tang et al. \shortcite{Tang2020HBM} developed a method for designing compliant mechanical systems with periodic large-amplitude oscillations.  
Also targeting the large-deformation regime, D{\"u}nser et al. \shortcite{Duenser2022} proposed a nonlinear extension to linear eigenmodes.
When applied to thin sheets, these nonlinear compliant modes tend to generate bending deformations initially, but as we show in our analysis, they ultimately induce stretching for larger deformations. By contrast, our approach produces large folding transformations with much lower elastic energy.

\paragraph{Rest-State Modifications \& Intrinsic Actuation}
The concept of driving elastic deformations through rest state modifications has been explored in the past \cite{Kondo05Directable,Schumacher12}. Several works have used volumetric blend shapes to drive rest poses for physics-based facial animation \cite{Kozlov17Enriching,Ichim17Phace}.
Similarly, intrinsic actuation via rest-state modifications has been leveraged for motion control of deformable characters \cite{Coros2012Deformable,Tan12Soft}.
Similar to these existing works, our approach builds on rest curvature modifications to drive folding transformations.
To the best of our knowledge, however, our work is the first to combine the concept of intrinsic actuation with modal analysis.

\paragraph{Shape Interpolation} 
Interpolating between a set of input shapes is a basic task in geometry processing \cite{Alexa00ARAP}. Since the body of literature is too large for an exhaustive review, we focus on methods that interpolate triangle meshes using shell-like deformation energies. The common approach in this context is to first convert shapes into differential coordinates---e.g., dihedral angles and edge lengths---that are invariant under translation and (often) rotation. Building on this basis, Fr{\"o}hlich et al.~\shortcite{frohlich2011example} linearly interpolated between differential coordinates of input shapes. The interpolated shape is obtained by minimizing a discrete shell energy \cite{grinspun2003discrete} with rest angles and edge lengths given by the interpolated differential coordinates. Similarly, our method also builds on linear interpolation in a space of nonlinear differential coordinates to which we refer as strain space. However, instead of interpolating between example shapes, we extrapolate a single shape using strain-space directions induced by linear eigenmodes.

Another closely related concept is geodesic interpolation in shape space \cite{Kilian07Geometric}. Heeren et al.~\shortcite{heeren2012time} obtained the shortest ge\-o\-de\-sics between pairs of shapes by minimizing the total dissipation of the discrete shell energy along the path. Similarly, Heeren and colleagues \shortcite{heeren2014exploring} compute geodesics by minimizing the path energy defined through a sequence of deformed shapes. In addition, several works leverage Principal Geodesic Analysis (PGA)  for capturing the variability of a set of shapes~\cite{freifeld2012lie,heeren2018principal, sassen2020nonlinear}. All the above methods require either a second shape or a tangent space direction, which must be specified by the user. Our approach automatically computes energetically efficient folding transformations from a single input shape using the eigenstructure of its Hessian.

\section{Method}
Our approach leverages the mechanics of thin sheet materials to compute folding transformations in an automated manner. We begin with a brief review of linear modes before we proceed to our nonlinear strain-space modes.

\subsection{Eigenmodes}
\label{sec:modalAnalysis}

We use a standard discrete shell model \cite{gingold2004discrete,grinspun2003discrete} to describe the mechanics of thin sheet materials (see 
Supp. 1). Using this model, we compute energetically optimal folding transformations that minimize the mechanical work required to achieve deformations of prescribed magnitudes. Let $\bX$ and $\bx$ denote vertex positions for undeformed and deformed states, respectively. Furthermore, let $\bff=-\nabla_\bx W(\bX,\bx)$ denote the vector of forces and $\bH=\nabla^2 W$ the corresponding Hessian arising from the discrete total energy $W$. This energy includes internal elastic energy and external penalty energies for fixing vertices to exclude rigid transformations. For deformations around the rest state, a second-order expansion of the elastic energy yields
\begin{equation}
	W(\bX+\bu) = W(\bX)-\bff(\bX)^\tran\bu + \frac12\bu^\tran\bH\bu + O(\bu^3) \approx \frac12\bu^\tran\bH\bu \ ,
\label{eq:quadraticEnergy}	
\end{equation}
where we used the fact that both the energy and its gradient vanish at the reference configuration. Finding displacements that minimize the energy for given deformation budgets leads to a constrained optimization problem,
\begin{equation}
    \min \frac12 \bu^\tran\bH\bu \qquad \text{s.t.} \quad \frac12\bu^\tran\bM\bu=b^2 \ ,
\end{equation}
where $\bM$ is the mass matrix and $b$ is a given displacement magnitude. The Lagrangian for this problem is
\begin{equation}
    \mathcal{L}(\bu,\lambda)= \frac12 \bu^\tran\bH\bu +\lambda(\frac{1}{2}\bu^\tran\bM\bu-b^2) \ ,
\end{equation}
where $\lambda$ is a Lagrange multiplier.
The first-order optimality conditions of this problem require that
\begin{equation}
\label{eq:optEigValue}
    \bH\bu +\lambda\bM\bu = 0 \ .
\end{equation}
It is evident from this expression that \textit{solutions to the above optimization problem are collinear with generalized eigenvectors $\be_i$ of the Hessian} $\bH$. In particular, the solution corresponding to the smallest eigenvalue $\lambda_0$ minimizes the mechanical work required to generate a displacement of magnitude $b$. Solutions for other eigenvalues $\lambda_i$ minimize work in the subspace orthogonal to the $i-1$ lower-order eigenvectors.

\begin{figure*}[t]
    \centering
    \includegraphics[width=0.9\linewidth]{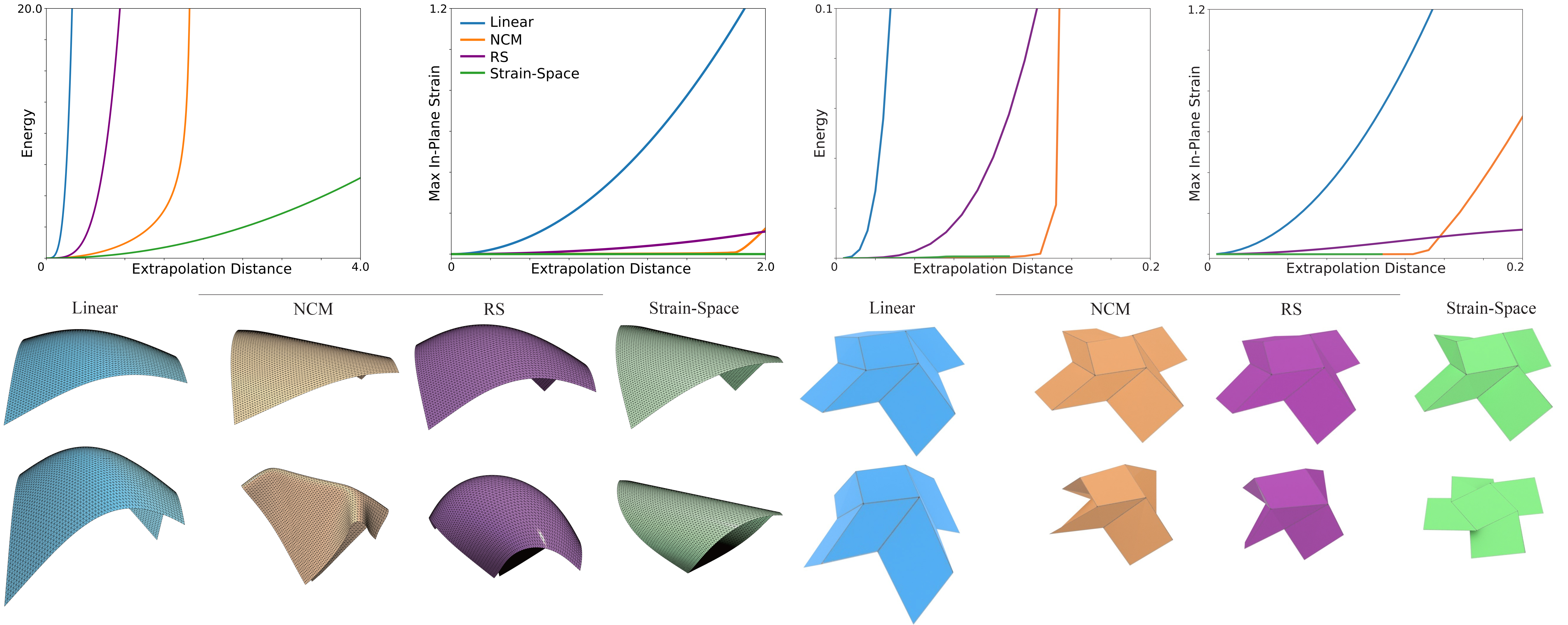}
    
    \caption{Comparing internal energy and maximum in-plane strain along the first non-rigid mode computed using linear eigenmodes, nonlinear compliant modes (NCM), rotation-strain modes (RSM), and our strain-space modes.
    }
    \label{fig:Energy}
\end{figure*}

These considerations suggest that generalized eigenvectors of the energy Hessian are promising directions for folding transformations, i.e., large-magnitude displacements that require little mechanical work. Indeed, when analyzing eigenvectors for the discrete shell energies described in 
Supp. 1, we see that low-order eigenvectors are normal to the surface for flat sheets. Since normal displacements do not generate stretching to first order, these eigenvectors correspond to bending deformations, which are energetically favorable.
However, simply following these eigenvectors along linear trajectories,
\begin{equation}
\Psi_i^\mathrm{LM}(t) = \bX+t\be_i \ ,
\end{equation}
does not lead to physically meaningful results: although linear modes produce bending initially, they tend to induce stretching for larger displacements, resulting in a steep increase in energy (see Fig. \ref{fig:Energy}). %
This behavior is explained by the fact that the discrete shell energy is highly nonlinear and the quadratic approximation (\ref{eq:quadraticEnergy}), from which eigenvectors are computed, is only valid for small displacements.

Many extensions of linear eigenmodes to the finite-deformation range have been explored in the past. We experimented with the two variants that we believe to be the most promising options. However, as our analysis in Sec. \ref{sec:results} shows, neither of them are able to produce the low-energy transformations that we seek to achieve. For this reason, we develop a new formulation that we describe next.

\subsection{Strain-Space Modes}

Linear modes ultimately suffer from a lack of rotational invariance: a linear trajectory in position-space cannot generate pure bending deformations. Our formulation, instead, leverages the rotational invariance of discrete shell energies. By modulating per-element rest curvatures, we can draw the equilibrium configuration of the thin sheet towards desired states. Let $\restcurvature=(\kappa_0,\ldots,\kappa_m)^\tran$ denote the vector of per element rest curvatures, which can either be per-edge dihedral angles $\kappa_i=\theta_i$~\cite{grinspun2003discrete} or per-triangle shape operators $\kappa_i=\bS_i$~\cite{gingold2004discrete}, depending on the type of bending element used (see Supp. 1). Let $W^\mathcal{M}(\bX, \bx)$ and $W^\mathcal{B}(\bX,\restcurvature,\bx)$ denote the discrete membrane and bending energies, respectively. Minimizing elastic energy $W(\bX,\restcurvature, \bx) = W^\mathcal{M}(\bX, \bx) + W^\mathcal{B}(\bX,\restcurvature,\bx)$ for given rest curvatures $\restcurvature$ will produce the deformed configuration $\bx^*$ that best approximates the target curvatures, 
\begin{equation}
\label{eq:curvaturesToPositions}
    \bx^*=\argmin_\bx W(\bX,\restcurvature,\bx) \ .
\end{equation}
We note that the minimizer will have zero energy if and only if the target curvatures are \textit{compatible} in the sense that they can be obtained without stretching the sheet. Otherwise, the minimizer will balance in-plane deformation and deviation from target angles in a material-aware manner.

Clearly, different target curvatures will lead to different equilibrium shapes. Since linear modes are energetically optimal for small displacements, we choose the per-element curvatures induced by linear modes at the rest configuration as target curvatures. To this end, we ask that the initial tangent of the system's trajectory $\bx(t)$ should coincide with the eigenvector $\be_i$, i.e., $\dot\bx(0)=\be_i$. The position-space velocity induces a rate of change in curvatures according to
\begin{equation}
    \dot\curvature = \frac{\partial \curvature}{\partial \bx}\frac{\partial \bx}{\partial t}
    =\bJ\be_i \ ,
\end{equation}
where the Jacobian $\bJ$ maps position-space directions into corresponding strain-space directions. Following the modal directions $\bJ\be_i$ defines linear strain-space trajectories $\restcurvature(t) = \restcurvature(0)+t\bJ\be_i$. 
We can reconstruct points on these strain-space trajectories into position-space configurations by using them as rest curvatures in (\ref{eq:curvaturesToPositions}). The corresponding nonlinear position-space trajectories are  obtained by solving minimization problems of the form
\begin{equation}
\label{eq:StrainSpaceModesContinuuous}
    \Psi^\mathrm{SSM}_i(t)=\argmin_\bx W(\bX,\restcurvature(t),\bx) \ .
\end{equation}
The strain-space modes defined in this way build on changes in rest curvature to drive shape transformation. This \textit{intrinsic} actuation mechanism is rotationally invariant by construction. As we show in Sec. \ref{sec:results}, this property makes strain-space modes ideal for exploring folding transformations of thin sheet materials.

\paragraph{Multi-Dimensional Extension}
Just as their linear counterparts, one-dimensional strain-space modes can be extended to multi-di\-men\-sion\-al nonlinear subspaces. To this end, we generalize the definition of rest curvatures,
\begin{equation}
\restcurvature(\bc) = \restcurvature(0)+\sum_jc_j\bJ\be_j \ ,
\end{equation}
where $\bc=(c_1, \ldots, c_m)$ is a vector of modal coefficients. The above definition in (\ref{eq:StrainSpaceModesContinuuous}) gives rise to a nonlinear subspace,
\begin{equation}
\label{eq:multiDimSubspace}
    \Phi^\mathrm{SSM}(\bc)=\argmin_\bx W(\bX,\restcurvature(\bc),\bx) \ .
\end{equation}
In Sec. \ref{sec:results} we illustrate the potential of this nonlinear subspace on an inverse folding example.
Finally, we note that the above definition
includes blended modes, i.e., generalized one-dimensional strain-space modes,
\begin{equation}
\label{eq:multiDimSSM}
    \Psi^\mathrm{SSM}(\hat{\bc},t)=\Phi^\mathrm{SSM}(t\hat{\bc}) \ ,
\end{equation}
where $\hat{\bc}$ is a unit-length vector of modal coefficients.

\paragraph{Implementation}
We compute discrete strain-space modes $\Psi^\mathrm{SSM}_{i,j}$ by minimizing (\ref{eq:StrainSpaceModesContinuuous}) with rest curvatures $\restcurvature=\Delta t\bJ\be_i$ ramping up according to the step size $\Delta t$.
We solve these unconstrained minimization problems using a standard Newton's method with adaptive regularization and line search.
To incorporate self-collision handling, we add the barrier potential 
proposed by Li et al.~\shortcite{li2020incremental}. To remove translational and rotational null spaces, we furthermore add quadratic penalties 
to fix selected vertices.
Since the resulting minimization problems are nonlinear and non-convex, they generally have multiple local minima. To avoid jumping between minima when computing the discrete trajectory, we initialize the Newton solve for a given next state $\Psi^\mathrm{SSM}_{i,j+1}$ with the previous state $\Psi^\mathrm{SSM}_{i,j}$. We found that this simple approach leads to smooth trajectories for all examples that we considered in this work.

\section{Results}
\label{sec:results}
We evaluate our method on a diverse set of examples and compare folding patterns generated through simulation with corresponding physical prototypes. We start by comparing strain-space modes to other alternatives. We then explore periodic patterns and demonstrate multi-dimensional strain-space modes on an inverse folding example. Unless otherwise specified, we use hinge-based bending elements \cite{grinspun2003discrete} as the default shell model for all experiments.

\subsection{Comparing Modes}
As a first experiment that allows for both quantitative and qualitative evaluation, we compare the lowest-energy mode produced by different methods for a square sheet with side lengths of $20cm$. We set Young's modulus to $2.9GPa$, Poisson's ratio to $0.3$, and use sheets of  $1mm$ thickness.
As can be seen from the energy plot shown on the left of Fig. \ref{fig:Energy}, linear modes induce stretching even for small deformations, leading to a steep energy profile. The nonlinear compliant modes by D{\"u}nser et al. \shortcite{Duenser2022} perform much better in comparison, showing a slower increase in energy for small to moderate deformations. However, as the diagonal tips of the sheet approach each other, the external actuation force aligns with the sheet's tangent space. This alignment induces stretching, which leads to a sharp increase in energy.
The reason for this behavior is that nonlinear compliant modes rely on  extrinsic actuation using an external force collinear with the linear mode. While a force with constant direction can create pure bending initially, it will induce stretching eventually. 
In contrast, our strain-space modes build on an intrinsic actuation method that modulates rest curvatures of discrete shell elements. The resulting actuation forces remain (approximately) normal to the surface even for large deformations. As a result, our strain space modes produce large bending deformations with virtually zero stretching and, consequently, a much slower increase in energy.
We furthermore experimented with another nonlinear extension of eigenmodes, the rotation-strain modes by Huang et al. \shortcite{Huang2011Interactive}. As can be seen from Fig. \ref{fig:Energy}, the lowest-order rotation-strain mode initially shows a slower increase in energy than its linear counterpart. However, it clearly exhibits double curvature, which translates into nonzero in-plane deformation and, consequently, a steep increase in energy beyond the small displacement regime. We conjecture that the least-squares geometry reconstruction step, which is at the heart of rotation-strain modes, does not distinguish between bending and stretching, leading to configurations that exhibit physically unrealistic levels of in-plane deformation.

As a second basic example, we compare the four different mode types on a simple origami design\footnote{https://origamisimulator.org/} consisting of a square sheet divided into 3$\times$3 sub-squares, each with a side length of $7.2cm$. The creased sheet exhibits a single degree of freedom, corresponding to a nonlinear twist motion of the center square.
To emulate the effect of pre-scoring, we reduce the bending stiffness for crease edges by a factor of 10. As shown on the right of Fig. \ref{fig:Energy}, only our method successfully reproduces the twisting behavior; see also the accompanying video. Additionally, the energy and in-plane strain plots show that our method achieves the final twisted state with low energy and small maximum in-plane strain.

These two examples underscore the observation that, unlike existing nonlinear modes, our strain-space modes give rise to energetically efficient folding transformations with very low levels of stretching. In the following examples, we further explore strain-space modes and their applications.

\subsection{Exploring Strain-Space Modes}
\begin{figure}
    \centering
    \begin{tabular}{ccc}
    Mode 25 & Mode 30 & Mode 35\\
    \includegraphics[width=0.26\linewidth]{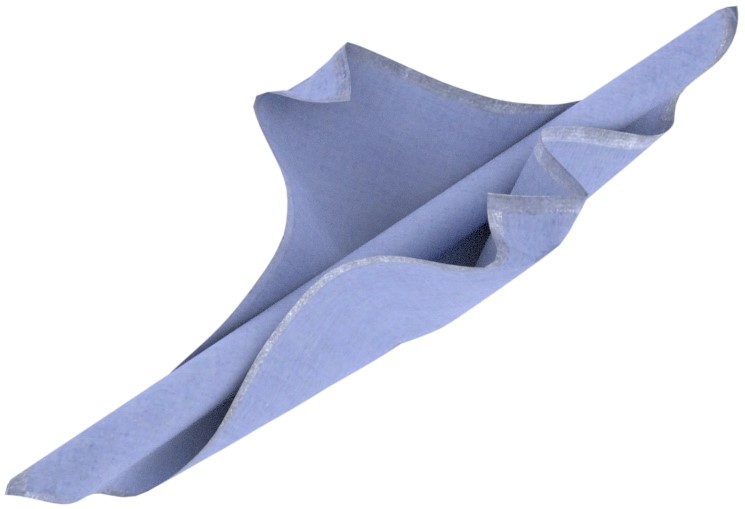} & \includegraphics[width=0.26\linewidth]{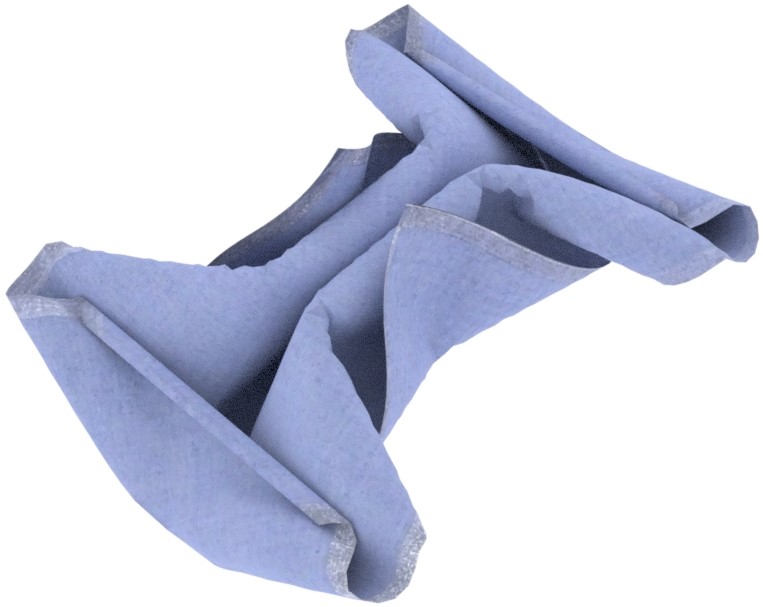} & \includegraphics[width=0.26\linewidth]{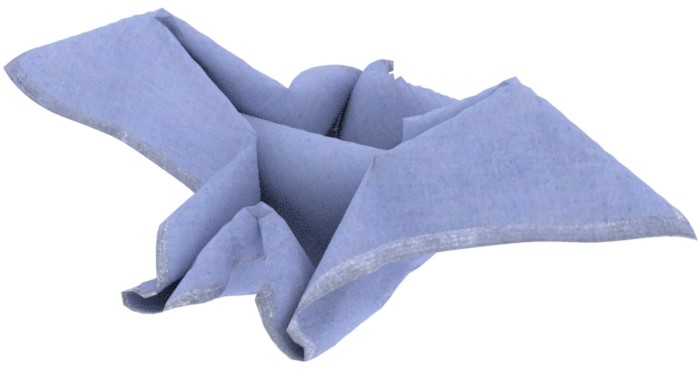} \\
    Mode 43 & Mode 50 & Mode 57\\
    \includegraphics[width=0.26\linewidth]{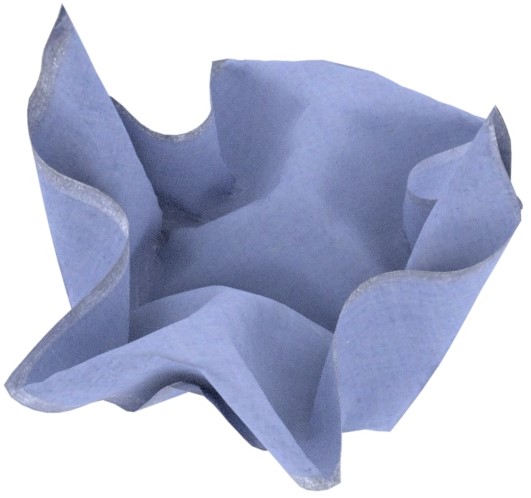} & \includegraphics[width=0.26\linewidth]{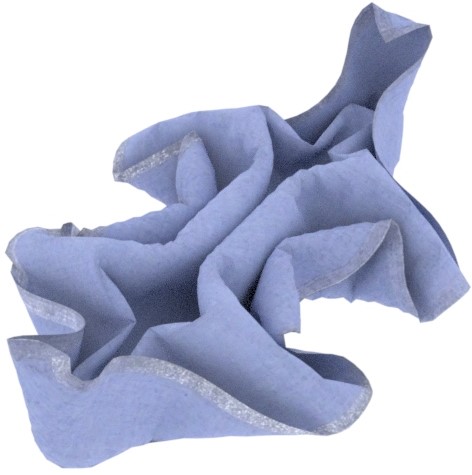} & \includegraphics[width=0.26\linewidth]{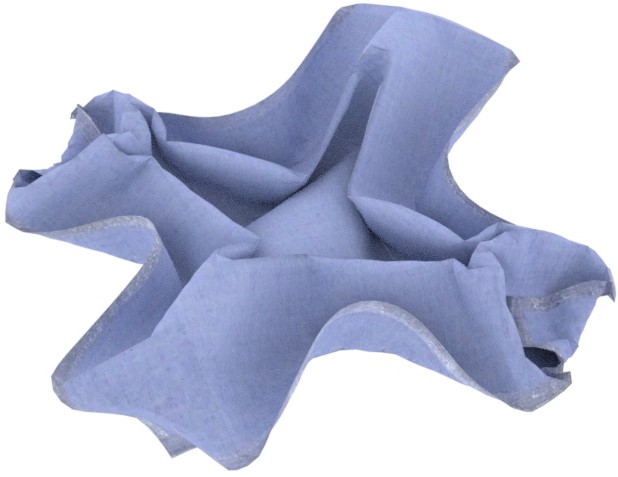}
    \end{tabular}
    \caption{Modal folding patterns for a square sheet.}
    \label{fig:squareSheetModes}
\end{figure}

Strain-space modes can generate diverse and interesting folding transformations for a large variety of input shapes and materials. Fig. \ref{fig:squareSheetModes} shows a selection of strain-space modes for the square sheet used in the initial example of Fig. \ref{fig:Energy}. Please see the accompanying video for an exhaustive tour through the modal shapes up to mode index 99.
It can generally be noted that the folding transformations computed with our method lead to large bending deformations with negligible stretching. Strain-space modes furthermore reflect the symmetries of their linear counterparts, but the nonlinearity of our approach gives rise to a rich space of intricately folded shapes.

To validate that the folding transformations generated by our method are physically feasible, we build physical prototypes using thin cotton sheets. To this end, we select and attach several pairs of points in close proximity such that the fabric maintains the desired folded state at equilibrium.
We developed a simple interactive tool for this purpose in which users select pairs of vertices that are attached using zero-length springs. The static solver then provides feedback on the deviation between equilibrium and target states.
We repeat the procedure until a desired shape accuracy is obtained. See the accompanying video for a demonstration of this process. We create physical prototypes for folded configurations using adhesive to connect the pairs of points selected by the user.
As can be seen from the example shown in Fig. \ref{fig:teaser}, our prototypes successfully reproduce the folded shapes predicted in simulation.

\begin{figure}
    \centering
    \begin{tabular}{ccc}
     Mode 16 & Mode 20& Mode 21 \\ \includegraphics[width=2.35cm]{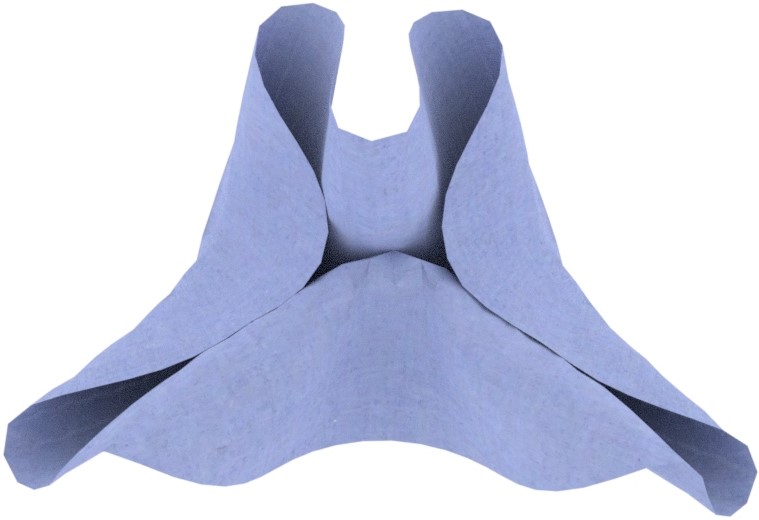} & \includegraphics[width=2.35cm]{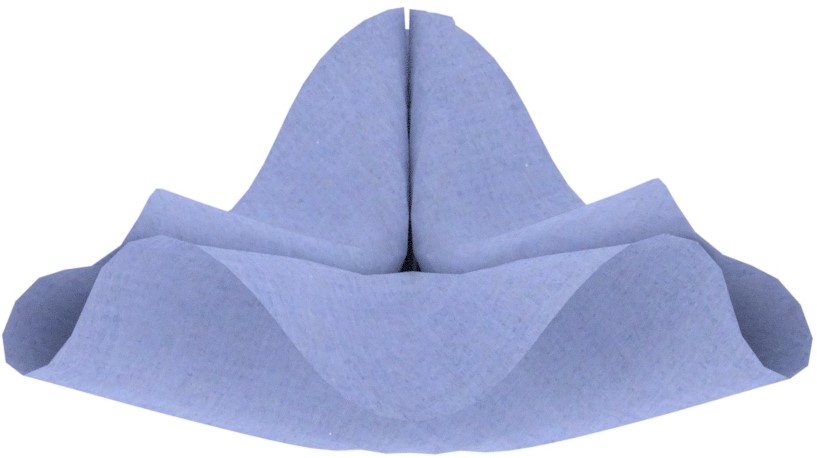} & \includegraphics[width=2.35cm]{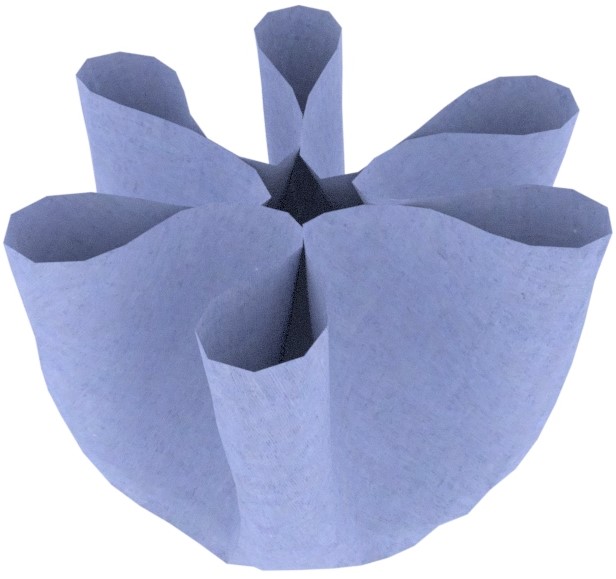} \\
      \\
     Mode 24 & Mode 25 & Mode 30 \\ 
     \includegraphics[width=2.35cm]{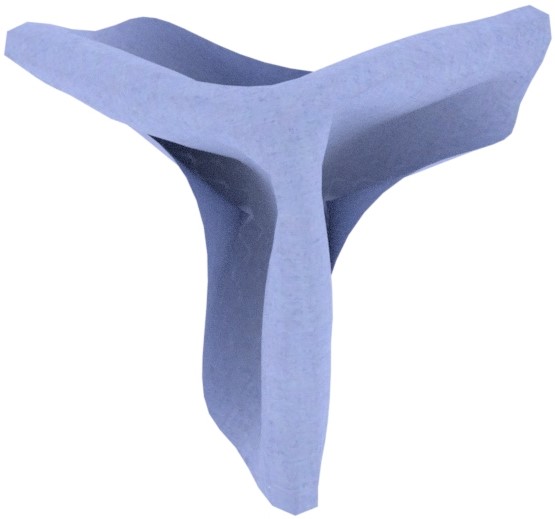} & \includegraphics[width=2.35cm]{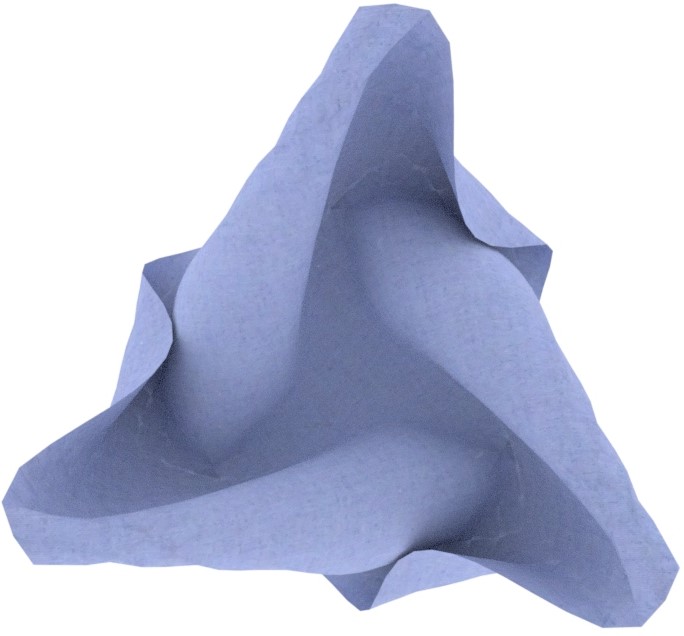} & \includegraphics[width=2.35cm]{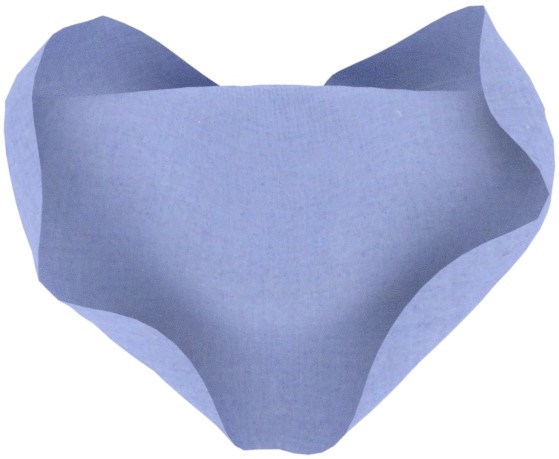} \\
    \end{tabular}
    \caption{Symmetric modal folding patterns for a disc-shaped sheet.}
    \label{fig:Disk}
\end{figure}

The shape of the input mesh has a significant impact on the resulting linear modes, and this effect translates directly to strain-space modes. For example, compared to the square sheet, the higher symmetry of the disc-shaped sheet shown in Fig. \ref{fig:Disk} leads to a different and more symmetric set of modal shapes. 

\begin{figure}
    \centering
    \begin{tabular}{ccc}
       \rotatebox{90}{Rest State} &
        \includegraphics[width=3.4cm]{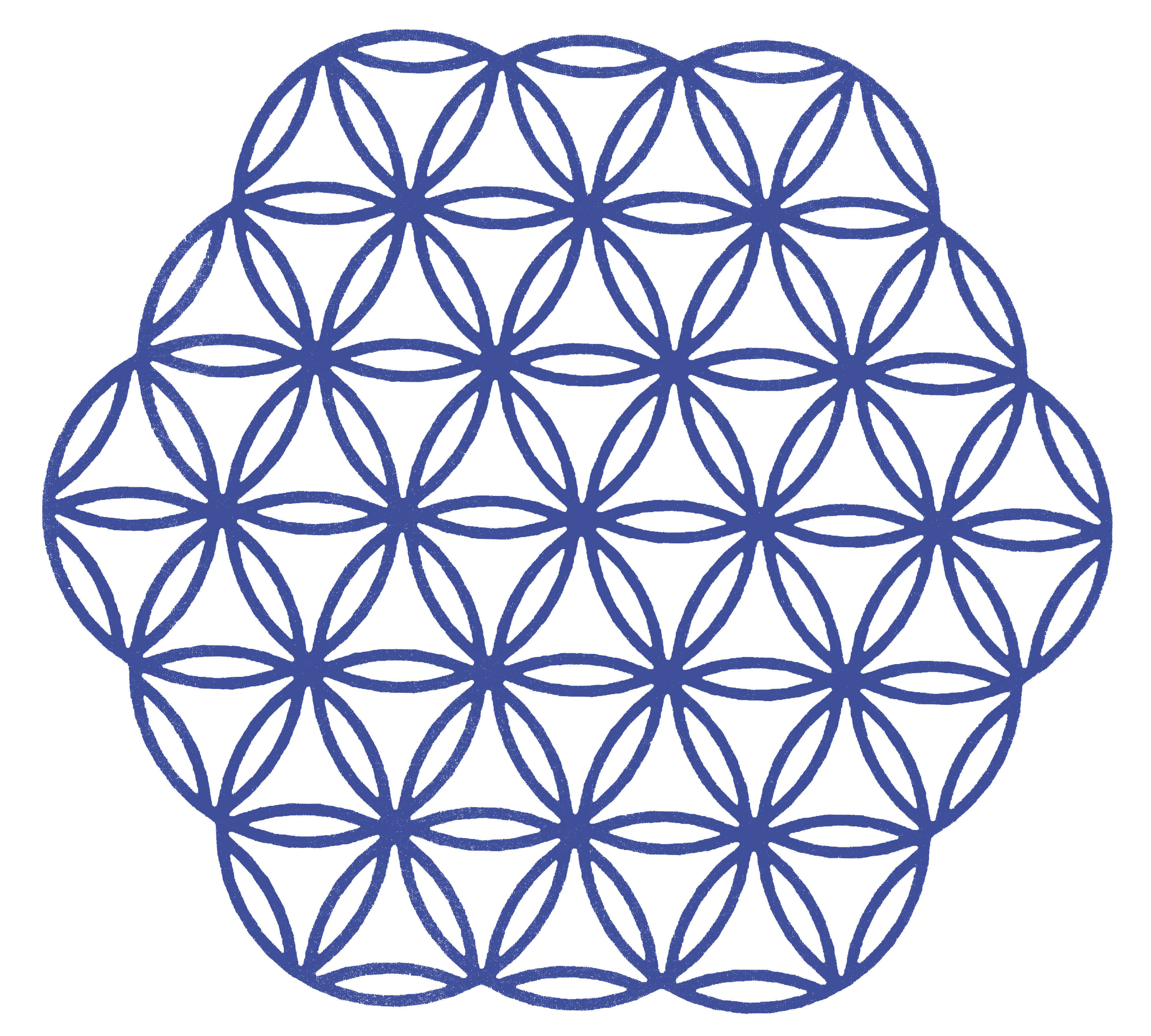} & \includegraphics[width=3.4cm]{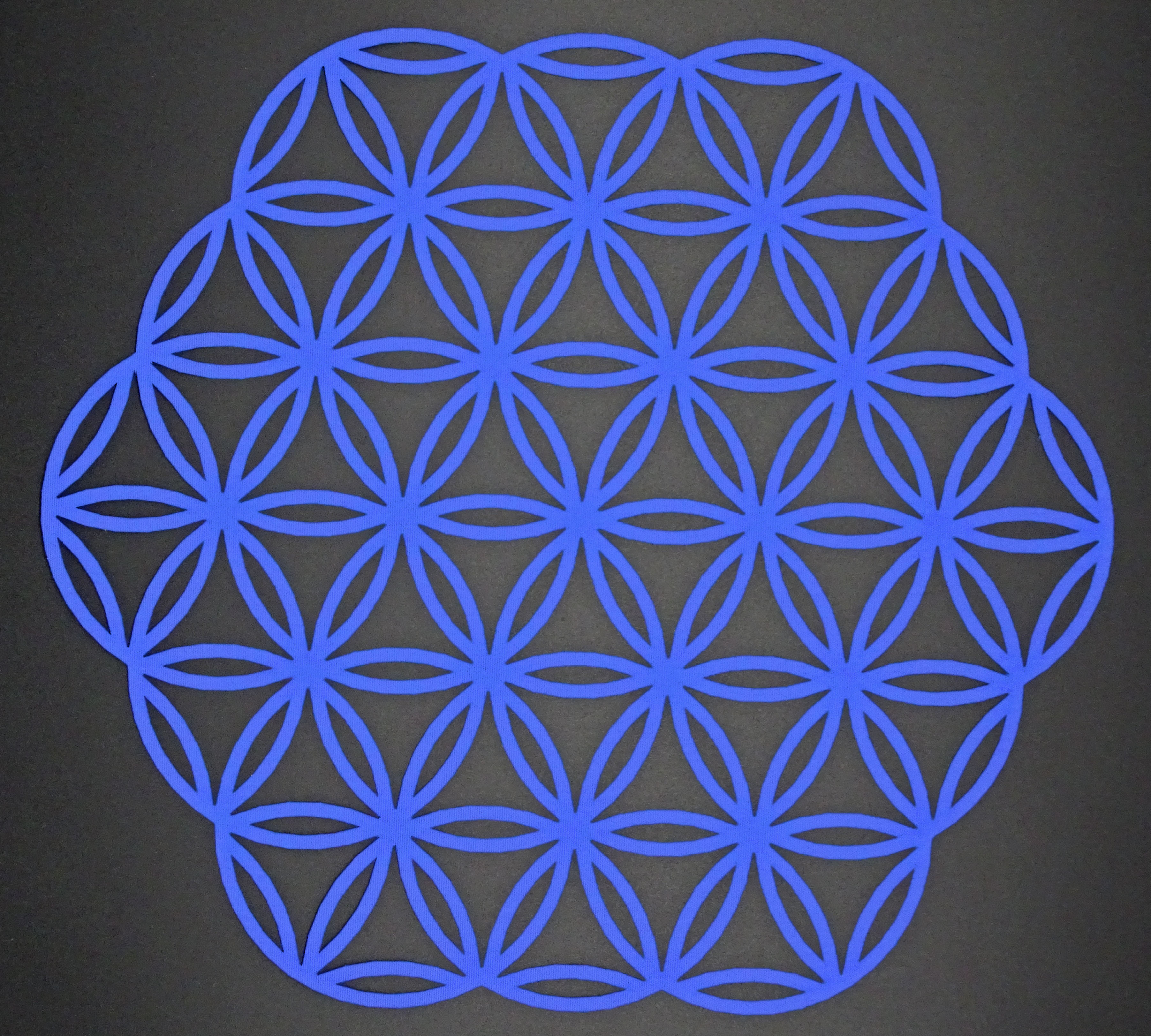}\\
       \rotatebox{90}{Mode 9} &
        \includegraphics[width=3.3cm]{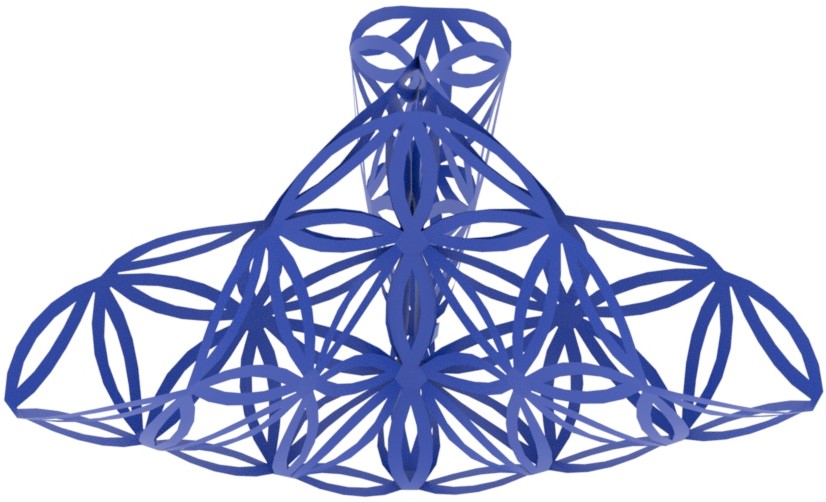} &    \includegraphics[width=3.4cm]{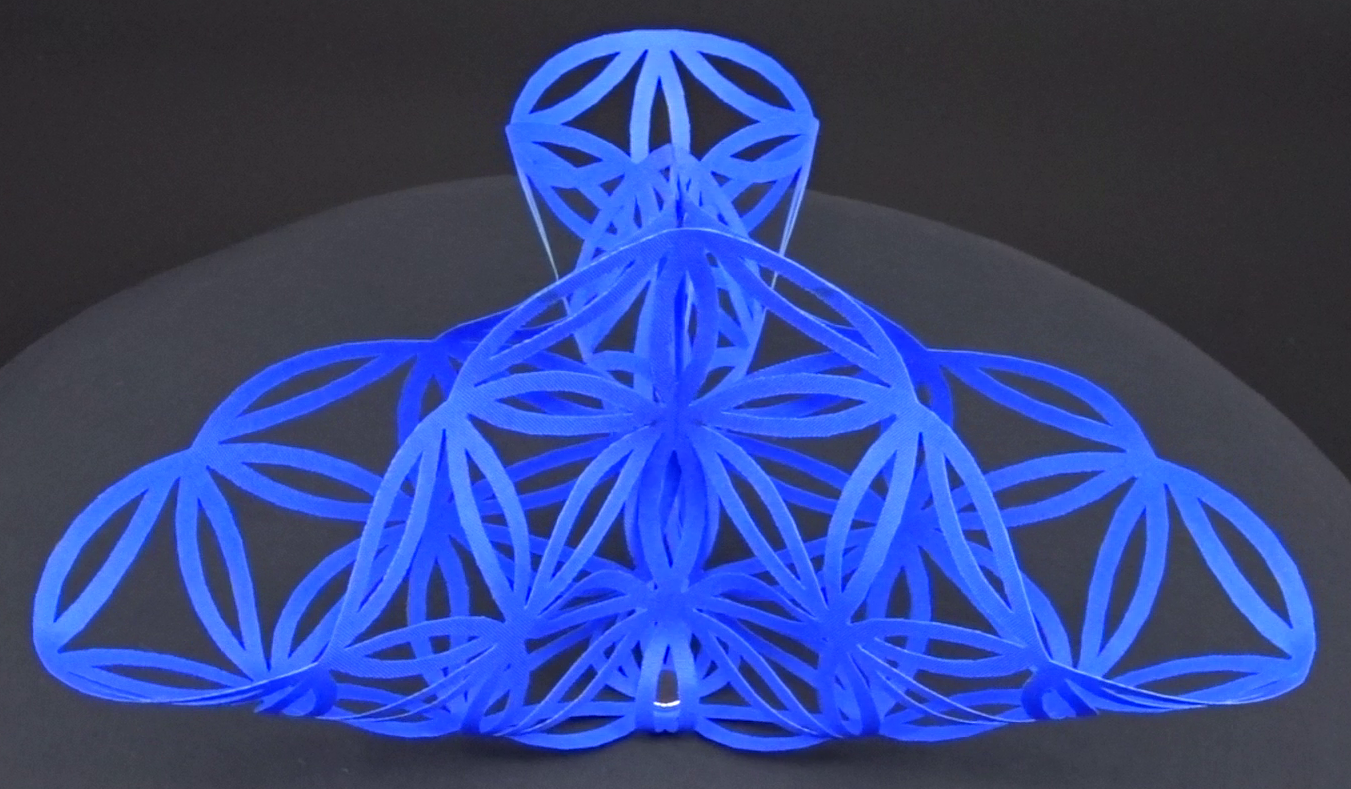}\\
       \rotatebox{90}{Mode 13} &
        \includegraphics[width=3.8cm]{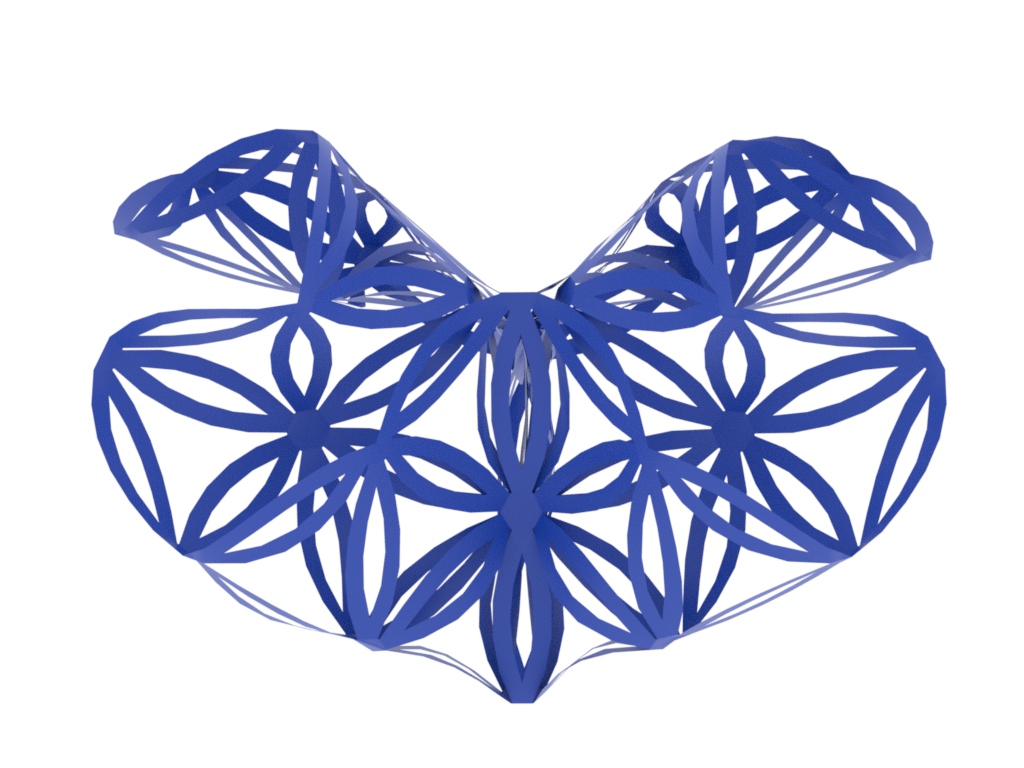} &    \includegraphics[width=3.4cm]{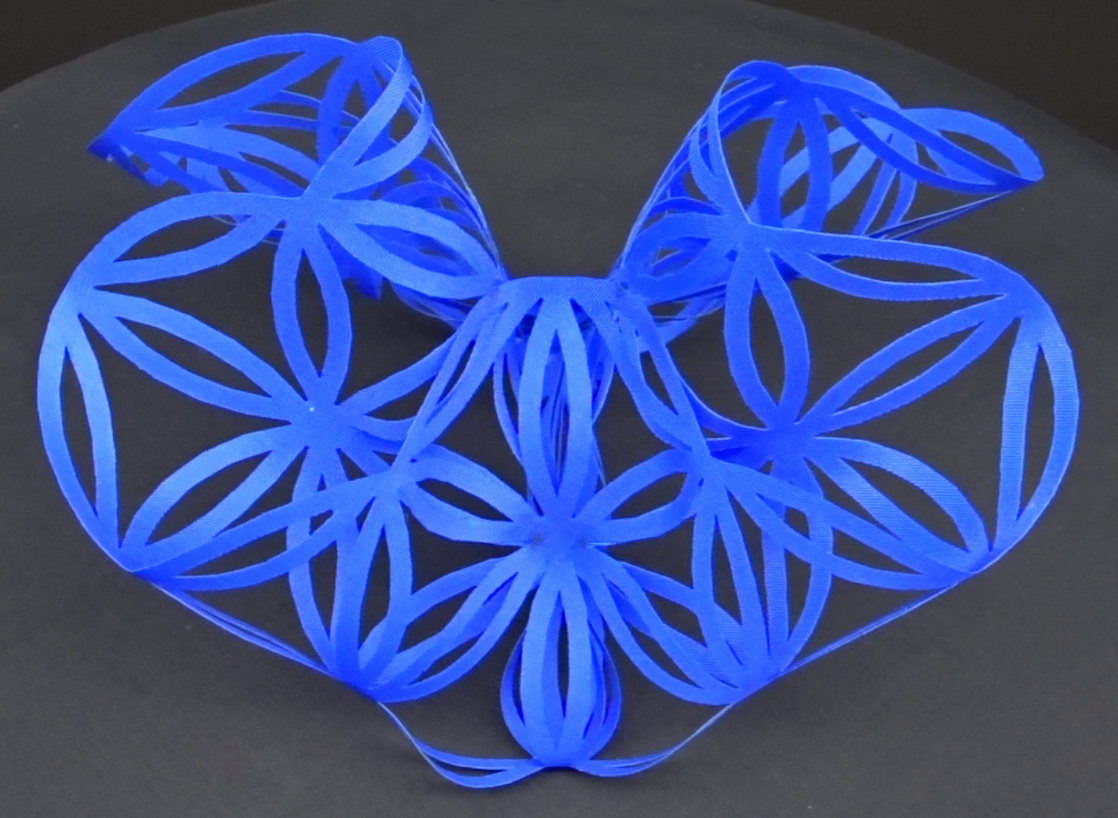}\\
       \rotatebox{90}{Mode 24} &
        \includegraphics[width=3.3cm]{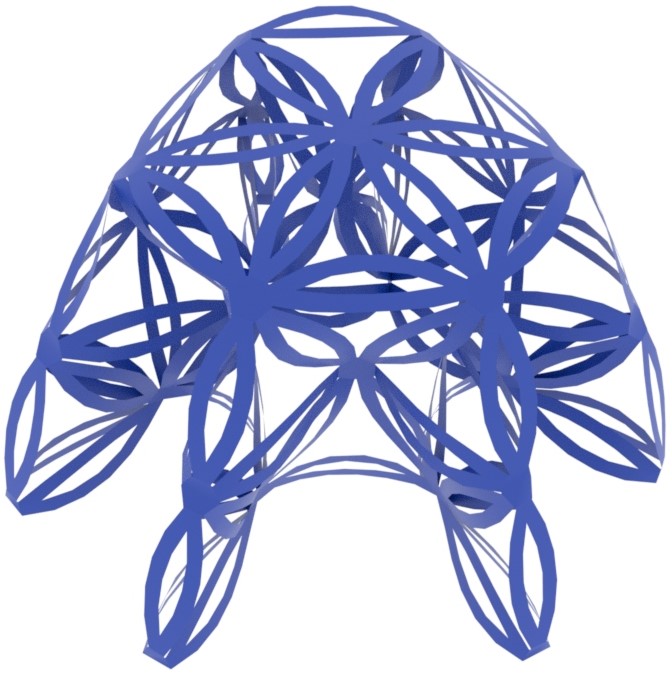} &    \includegraphics[width=3.4cm]{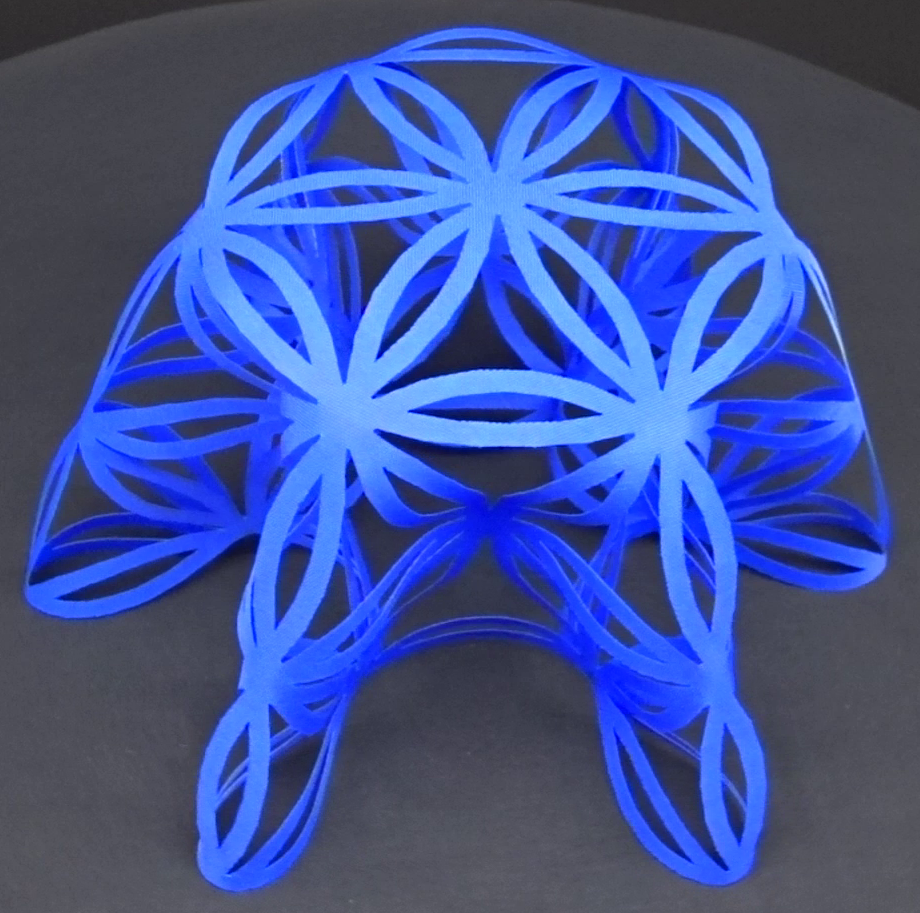}
    \end{tabular}
    
    \caption{Folding patterns for a thin sheet cut into the pattern of a life flower.}
    \label{fig:lifeFlower}
\end{figure}

In addition to input geometry, modal shapes are also affected by internal structure. For example, when removing material from a solid sheet, our method can leverage the void regions to induce in-plane deformations that lead to more efficient transformations. 
To illustrate this effect, we use a thin sheet cut into the shape of a so called \textit{life flower} pattern as shown in Fig. \ref{fig:lifeFlower}. 
Prototypes are fabricated using thick nylon fabric that is laser cut, folded, and fixed into position using instant glue. The physical shapes obtained in this way closely match the results obtained from the simulations.

\subsection{Periodic Folding Patterns}
Periodicity plays a central role in many folding patterns, offering both aesthetic appeal and functional advantages. Examples include smocking patterns for cloth design, self-folding periodic origami, and seamlessly tilable panels for architectural facades. 
In the following, we explore two types of periodic folding patterns based on translation and reflection tilings of the plane.

\begin{figure}
    \centering
    \includegraphics[width=0.95\linewidth]{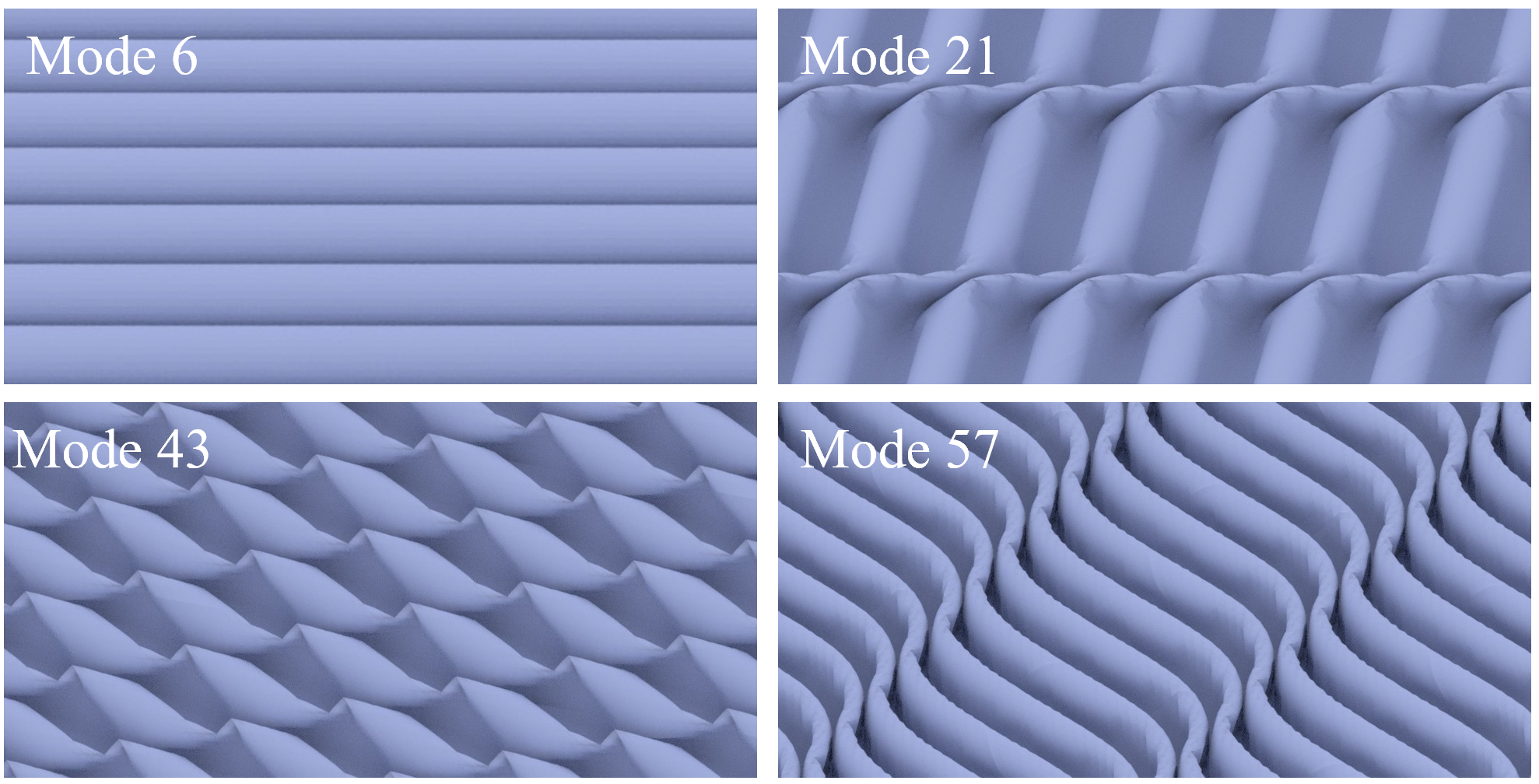}
    \caption{Strain-space modes for a square sheet subject to periodic boundary conditions.}
    \label{fig:PBC}
\end{figure}

\paragraph{Translation Tilings.} 
A translation tiling of the plane with a single shape means that pairs of opposite tile boundaries---left and right, top and bottom---have to match without gaps and overlaps. To enforce this periodicity condition, we ask that vertices $\bx_{ik}$ on boundary $i$ should have the same displacement $\bT_{ij}$ as their counterparts $\bx_{jk}$ on the opposite boundary $j$, i.e, $\bx_{jk}=\bx_{ik}+\bT_{ij}$ for all boundary vertices $k$.
Following previous work \cite{Schumacher18,Tang23Beyond,Li23Neural}, we implement these constraints by eliminating degrees of freedom from one side of each boundary pair and adding  the constant displacement between the two sides as additional variables. This transformation yields an unconstrained minimization problem whose Hessian gives rise to a set of eigenvectors that automatically satisfy the periodicity conditions. We convert these eigenvectors into strain-space modes as before---see Fig.~\ref{fig:PBC} for results.

\begin{figure}
    \centering
         \includegraphics[height=3.8cm]{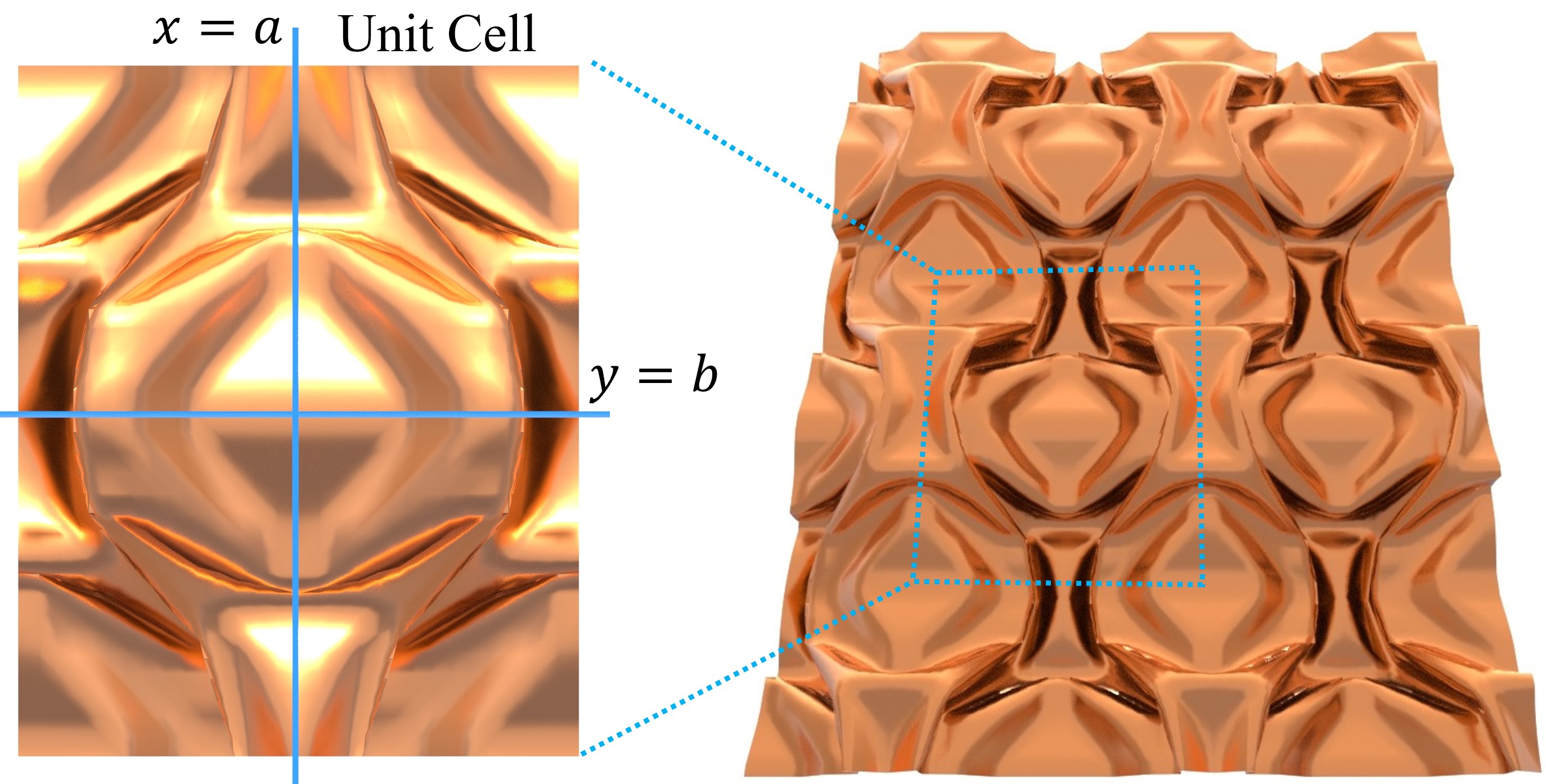}
    
    \caption{Folding results for a square sheet tiled with reflections at mode 23. We reflect the unit cell (the top-right tiling) to other tilings according to reflection planes $x=a$ and $y=b$. }
    \label{fig:AdditionalReflection}
\end{figure}

\paragraph{Reflection Tilings.}
Whereas translation tilings ask that a given boundary matches with its opposite side, reflection tilings require each boundary to tile with itself. To this end, all vertices of a given boundary must lie in a common reflection plane. We focus on flat rectangular tiles with axis-aligned $xz$- and $yz$- reflection planes. This setup leaves two degrees of freedom per boundary pair, i.e., $x=\pm a$ and $y=\pm b$,  where $a$ and $b$ denote the distance of the reflection plane from the origin of the patch; see Fig. \ref{fig:AdditionalReflection}. Removing all in-plane degrees of freedom for boundary vertices and adding the two variables $a$ and $b$, we obtain an unconstrained minimization problem, from which we  compute periodic modal tilings in analogy to the translation-only case.
Examples of periodic folding patterns with reflection boundaries are shown in Figs. \ref{fig:reflection} and \ref{fig:AdditionalReflection}.
We use a copper-like material with $0.01mm$ thickness, resulting in somewhat sharper creases compared to cotton. To validate our simulation results, we fabricate several prototypes using thin copper sheets. This process involves matching 3D-printed molds that are tightly pressed against the copper sheets to shape them into the desired pattern. 
It can be seen from the results shown in Fig.  \ref{fig:reflection} that our physical prototypes reproduce the folding patterns obtained in simulation without noticeable artifacts.

\subsection{Inverse Design}
\label{sec:inverseDesign_results}
\begin{figure}
    \centering
    \includegraphics[width=4.0cm]{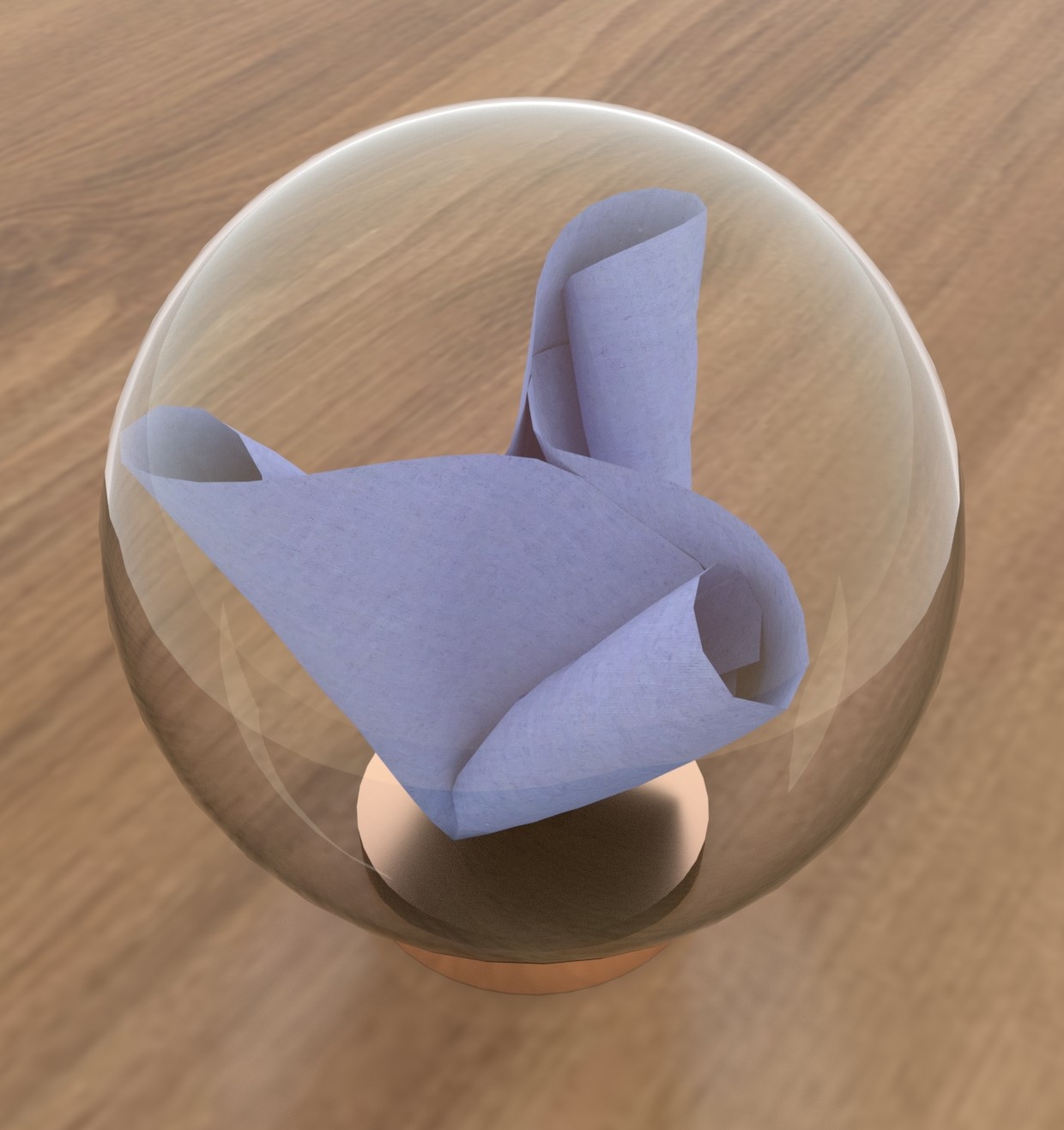}
    \includegraphics[width=4.0cm]{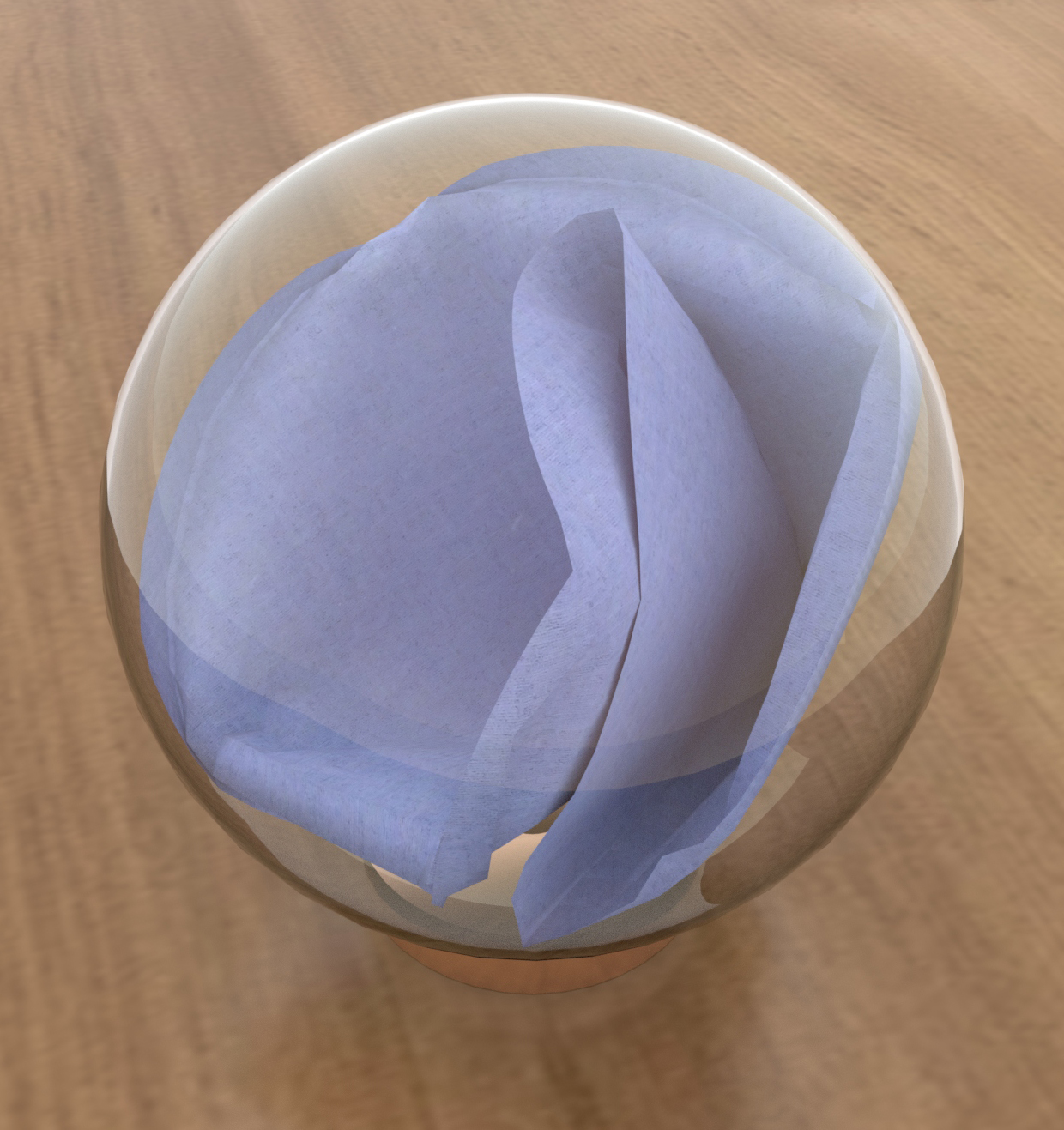}
    \caption{Inverse design for fitting a flat disc of radius $10cm$ into a sphere with a radius of $6cm$. Using the first 50 strain-space modes we obtain a folding pattern on the left. In comparison, we gradually decrease the radius of a sphere and obtain a crumpled folding on the right.}
    \label{fig:inverseDesign}
\end{figure}
The examples presented so far focused on forward exploration of one-dimensional modes. However, strain-space modes and their multi-dimensional extensions can also be used in inverse design problems. For the purpose of illustration, we consider an inverse folding problem where the goal is to find a folding transformation for a fabric disc of $10cm$ radius such that it fits into a sphere with radius of $6cm$. 
To this end, we define an objective function that penalizes mesh vertices outside the target sphere. We use the first 50 strain-space modes to define a nonlinear subspace as in (\ref{eq:multiDimSubspace}) and optimize for modal coefficients such as to minimize the objective function; see 
Supp. 2 for details. For comparison, we use IPC \cite{li2020incremental} to simulate the contact-induced compaction of the disc by gradually decreasing the radius of the sphere from $10.1cm$  to $6.0cm$ with a step size of $0.1cm$.  
As can be seen from the result shown in Fig. \ref{fig:inverseDesign}, inverse design with strain-space modes leads to a smooth and three-fold symmetric folding pattern. The contact-driven variant, in contrast, appears crumpled rather than folded.

\subsection{Additional Results and Experiments}
\paragraph{Performance.} We show the performance of our method by measuring the average time spent on computing each state using Newton's solver, considering 10 modes (from mode 6 to 15) and 10 states for each mode with an extrapolation step of 0.1. We test examples on a machine with an Intel Core i7-13700F 2.1GHz processor and 32 GB of RAM. Statistics are in Tab. \ref{tab:timePerformance}.

\begin{table}[t]
	\centering
	\caption{Performance of our method for the life flower, a disc sheet, and square sheets.}
	\begin{tabular}{ccccc}
		\toprule
		Example & life flower  & disc & square 40x40 & square 60x60 \\
        \midrule
        \#nodes & 1975& 817 & 1681 & 3721\\
        \#elements & 2262 & 1536 & 3200 & 7200\\
		Time[s] & 3.48 & 1.53 & 3.87 & 7.96\\
		\bottomrule
	\end{tabular}
	\label{tab:timePerformance}
\end{table}

In addition to the examples presented in this section, we evaluate the impact of mesh resolution and discretization on the resulting modal shapes in 
Supp. 4. We furthermore demonstrate in 
Supp. 3 that strain-space modes reflect the mechanical properties of the underlying material, including thickness-to-width ratio and isotropic vs. orthotropic bending stiffness.

\section{Conclusions}
We presented a new method to automatically discover folding patterns for thin sheet materials. The technical core of our approach is formed by strain-space modes that track the nonlinear bending strains induced by linear eigenmodes. We showed that strain-space modes lead to energetically favorable deformations compared to existing alternatives. Our examples reveal a rich space of novel folding patterns that can be realized as physical surfaces made from fabric, paper, or sheet metal.

\subsection{Limitations and Future Work}

\paragraph{Stretching.} Although strain-space modes are driven through curvatures, they are not automatically developable: per-element bending strains can combine into non-zero Gaussian curvature at vertices, which in turn induces stretching.
Empirically, we observed that stretching is generally small (
Supp. 3, Fig. 1). Nevertheless, developing a formulation for modes that are isometric by design would be an interesting subject for future work.

\paragraph{Mesh Dependence.} Our method behaves consistently under refinement, i.e., subdividing a mesh of given topology will generally not lead to drastic changes in modal shapes (
Supp. 4, Fig. 3). By contrast, changing mesh topology for a given resolution can change individual strain-space modes substantially (
Supp. 4, Fig. 4). Energy-minimizing remeshing could be a promising strategy for mitigating this effect. 

\paragraph{Blending Modes.}
We selected modes that we found aesthetically pleasing, interesting, and diverse. We did not perform in-depth experimentation with mode blending (\ref{eq:multiDimSSM}) as the resulting modes tend to have lower symmetry. Nevertheless, our inverse design example suggests that mode blending can be an effective means of discovering efficient folding transformations for high-level objectives.

\paragraph{Sharp Creases.}
Our method creates folds whose smoothness depend on the underlying material. Thinner sheets will generally produce sharper folds (
Supp. 3, Fig. 1). Nevertheless, it would be interesting to directly generate creases through modal folding. One promising approach to this goal might be to replace the elastic energy with a version that eliminates resistance to bending for larger curvatures, thus promoting creases.

\begin{acks}
The authors thank the anonymous reviewers for their valuable feedback. This work was supported by the European Research Council (ERC) under the European Union’s Horizon 2020 research and innovation program (grant agreement No.866480), and the Swiss National Science Foundation through SNF project grant 200021\_200644.
\end{acks}

\bibliographystyle{ACM-Reference-Format}
\bibliography{reference}
\end{document}